# Kinetic Analysis of Illite Dehydroxylation Based on Differential Scanning Calorimetry

Tomáš Ondro[1, a)], Anton Trník[1, 2, b)] and Omar Al-Shantir[1, c)]

[1]Department of Physics, Constantine the Philosopher University in Nitra, A. Hlinku 1, 949 74 Nitra, Slovakia
[2]Department of Materials Engineering and Chemistry, Czech Technical University in Prague, Thákurova 7, 166 29 Prague, Czech Republic

[a)] Corresponding author: tomas.ondro@ukf.sk
[b)] atrnik@ukf.sk
[c)] omar.alshantir@ukf.sk

**Abstract.** The two-step dehydroxylation of illite is studied using the differential scanning calorimetry on powder samples with heating rates from 1 to 10 °C min$^{-1}$ in a dynamic argon atmosphere. The values of apparent activation energy and pre-exponential factor are calculated using the Kissinger method. The determined values of apparent activation energy and pre-exponential factor of *trans*-vacant layer dehydroxylation are $(227 \pm 6)$ kJ mol$^{-1}$ and $(2.87 \pm 0.09) \times 10^{13}$ min$^{-1}$, respectively. The results also show that this process can be characterized by 1D diffusion controlled reaction with instantaneous nucleation rate. For the *cis*-vacant layer dehydroxylation, the values of apparent activation energy and pre-exponential factor are $(242 \pm 10)$ kJ mol$^{-1}$ and $(5.37 \pm 0.23) \times 10^{12}$ min$^{-1}$, respectively. The value of Avrami peak factor for this step indicates diffusion controlled growth of the new phase with zero or decreasing nucleation rate.

## INTRODUCTION

Illite is a significant rock-forming mineral that is the main component of illitic clay [1]. Its structure consists of repeating tetrahedron-octahedron-tetrahedron (T-O-T) layers and the interlayer space is occupied mainly by potassium cations and a variable amount of water molecules [2].

During heating, in the temperature range of (450 – 800) °C, the dehydroxylation of illite occurs according to the reaction: $2(OH) \rightarrow H_2O(\uparrow) + Or$ (Or is residual oxygen) [2,3]. Then, the formed water molecules migrate throughout the thermally expanded tetrahedral rings into the interlayer region and finally out of the illite structure [4].

Thermogravimetric experiments show that this process proceeds in two overlapping steps. The most probable explanation seems to be that the individual steps correspond to dehydroxylation of *trans*-vacant (*tv*) and *cis*-vacant (*cv*) illite layers [5]. The values of apparent activation energy for *tv*- and *cv*-layer dehydroxylation show considerable variations. An overview of determined values of activation energy is shown in Tab. 1. For this reason, the further systematic study of these processes is required.

**TABLE 1.** Overview of published values of activation energy for *tv*- and *cv*-layer dehydroxylation.

| Analysis | *tv*-layer dehydroxylation kJ mol$^{-1}$ | *cv*-layer dehydroxylation kJ mol$^{-1}$ | References |
|---|---|---|---|
| X-Ray powder diffraction (XRPD) | 697 | 231 | [4] |
|  | 676 | 230 |  |
| Thermogravimetric analysis (TGA) | 123.8 | 109.4 | [6] |
| Thermodilatometric analysis (TDA) | 119 | 184 | [5] |

The aim of this study is the kinetic analysis of illite dehydroxylation. From the results of differential scanning calorimetry (DSC) the values of apparent activation energy and pre-exponential factor were calculated by the Kissinger method.

## MATERIAL AND METHODS

High-purity illitic clay mined in Füzérradvány location in North-Eastern Hungary [1] was used in this study. According to [7], it can be characterized as 1-M polytype with an ideal chemical formula

$$K_{0.78}Ca_{0.02}(Mg_{0.34}Al_{1.69}Fe^{III}_{0.02})[Si_{3.35}Al_{0.65}]O_{10}(OH)_2 \cdot nH_2O. \tag{1}$$

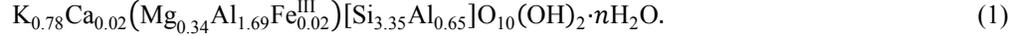

For the analysis, the raw illitic clay was crushed and milled in planetary ball mill Retsch PM 100 to obtain a powder with grains size less than 100 μm. The measurements were performed using Netzsch DSC 404 F3 Pegasus in dynamic Argon atmosphere with a flow rate of 40 ml min$^{-1}$ on samples with mass ~40 mg. The heating rates used in the experiments were 1, 3, 5, 7, and 10 °C min$^{-1}$.

## CALCULATION OF KINETIC PARAMETERS

For the parameterization of the illite dehydroxylation, the Kissinger method [8] was used. It is based on equation

$$\ln\left(\frac{\beta}{T_m^2}\right) = \ln\left[\frac{AR}{E_A}n(1-\alpha_m)^{n-1}\right] - \frac{E_A}{RT_m}, \tag{2}$$

where $\beta$ is the heating rate, $T_m$ is the temperature at which the process has a maximum rate, $A$ is the pre-exponential factor, $E_A$ is the apparent activation energy, $R$ is the universal gas constant, $\alpha_m$ corresponds to the value of the degree of conversion at temperature $T_m$, and $n$ is the kinetic exponent. The value of apparent activation energy can be calculated from the slope of the plot of the left-hand side of the Eq. (2) vs $T_m^{-1}$.

For the calculation of pre-exponential factor, the value of $n$ must be determined. It can be calculated as [8]

$$n = 1.26\, S^{1/2}, \tag{3}$$

where $S$ is the shape index. The $S$ is defined as the absolute ratio of the slope of the tangents to the peak at the inflection points. It can be expressed analytically as [8,9]

$$S = \frac{(\partial^2 N/\partial t^2)_{T_1}}{(\partial^2 N/\partial t^2)_{T_2}}, \tag{4}$$

where $N$ denotes the value of measured quantity and $t$ is time. $T_1$ and $T_2$ are frontal and terminal inflection points, respectively.

The mechanism of the reaction can be determined using the Avrami peak factor $n_A$ which is defined as

$$n_A = \frac{2.5\, T_m^2\, R}{w_{1/2}\, E_A}, \tag{5}$$

where $w_{1/2}$ is a peak's half-width [9].

### Peak separation

For the separation of overlapping peaks, the Fraser-Suzuki function [10] was used. It can be written as

$$y(x) = a_0 \exp\left[-\ln(2)\left[\frac{\ln\left(1 + 2a_1 \frac{x-T_m}{w_{1/2}}\right)}{a_1}\right]^2\right], \tag{6}$$

where $a_0$, $a_1$, $T_m$ and $w_{1/2}$ are the peak height, shape factor, position, and half-width, respectively. Values of these parameters can be found by using numerical optimization.

## RESULTS AND DISCUSSION

The results of DSC measurements with different heating rates in the temperature range of (30–1100) °C are shown in Fig. 1. The superposition of two Fraser-Suzuki functions (Eq. (6)) was fitted to the experimental data by numerical optimization, from which the values of $T_m$ and $w_{1/2}$ for $tv$- and $cv$-layers dehydroxylation were determined (see Table 2 and 3).

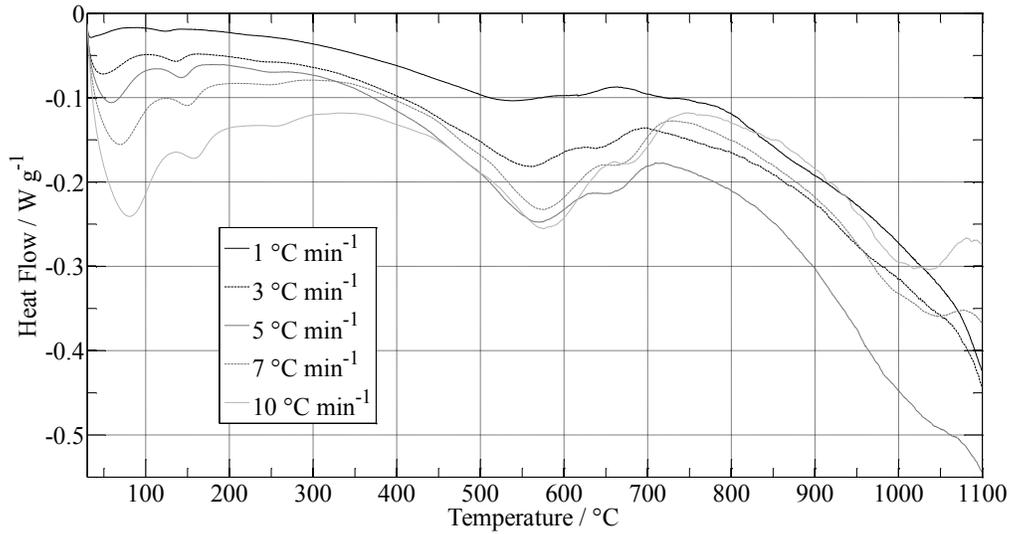

**FIGURE 1.** DSC curves of illite samples measured with different heating rates.

**TABLE 2.** Peak parameters of the $tv$-layer dehydroxylation.

| $\beta$ | $T_m$ | $w_{1/2}$ | $n_A$ | $T_i$ | $T_f$ | $T_1$ | $T_2$ | $S$ | $n$ |
| °C min$^{-1}$ | °C | °C | | °C | °C | °C | °C | | |
| --- | --- | --- | --- | --- | --- | --- | --- | --- | --- |
| 1 | 526.5 | 104.53 | 0.56 | 443.6 | 621.4 | 482.3 | 571.2 | 1.21 | 1.38 |
| 3 | 553.3 | 110.14 | 0.57 | 459.3 | 646.2 | 506.7 | 600.5 | 0.98 | 1.25 |
| 5 | 564.8 | 101.52 | 0.63 | 480.7 | 653.5 | 521.9 | 608.2 | 1.70 | 1.31 |
| 7 | 572.0 | 120.18 | 0.54 | 461.6 | 666.1 | 521.2 | 623.4 | 0.81 | 1.13 |
| 10 | 580.8 | 128.44 | 0.52 | 455.1 | 675.0 | 527.1 | 635.1 | 0.68 | 1.40 |

**TABLE 3.** Peak parameters of the $cv$-layer dehydroxylation.

| $\beta$ | $T_m$ | $w_{1/2}$ | $n_A$ | $T_i$ | $T_f$ | $T_1$ | $T_2$ | $S$ | $n$ |
| °C min$^{-1}$ | °C | °C | | °C | °C | °C | °C | | |
| --- | --- | --- | --- | --- | --- | --- | --- | --- | --- |
| 1 | 619.7 | 48.10 | 1.42 | 560.9 | 646.9 | 600.4 | 639.1 | 0.35 | 0.74 |
| 3 | 646.2 | 45.58 | 1.59 | 606.1 | 683.5 | 626.9 | 665.7 | 0.92 | 1.21 |
| 5 | 660.5 | 45.52 | 1.65 | 617.4 | 695.1 | 641.5 | 679.8 | 0.74 | 1.80 |
| 7 | 670.6 | 47.95 | 1.60 | 625.9 | 707.5 | 650.4 | 691.2 | 0.77 | 1.11 |
| 10 | 683.8 | 51.32 | 1.53 | 639.8 | 727.1 | 662.1 | 705.8 | 0.98 | 1.25 |

Figure 2 shows the Kissinger plot constructed from the determined values of $T_m$ for both reactions. This plot shows only one linear part, which indicates the independence of values of apparent activation energy of both steps on the heating rate. The value of apparent activation energy for $tv$-layer dehydroxylation is $(227 \pm 6)$ kJ mol$^{-1}$. This value is slightly higher than the values published in recent articles [5,6] and can be caused by the different experimental conditions. The value of Avrami peak factor $n_A$=0.56 ± 0.04 indicates 1D diffusion controlled reaction with instantaneous nucleation rate [4]. The same result was also published in [4].

The value of apparent activation energy for $cv$-layer dehydroxylation is $(242 \pm 10)$ kJ mol$^{-1}$ and the value of $n_A$ is 1.56 ± 0.08. The value of Avrami peak factor indicates diffusion controlled growth of the new phase with zero or decreasing nucleation rate [9]. This value is different from the results in [5], however, the diffusion controlled process was also found to be a dominant mechanism of this step. After determining the value of kinetic exponent $n$ the value of *A* can be calculated. The values of the pre-exponential factor for *tv*- and *cv*-layer dehydroxylation are $(2.87 \pm 0.09) \times 10^{13}$ min$^{-1}$ and $(5.37 \pm 0.23) \times 10^{12}$ min$^{-1}$, respectively.

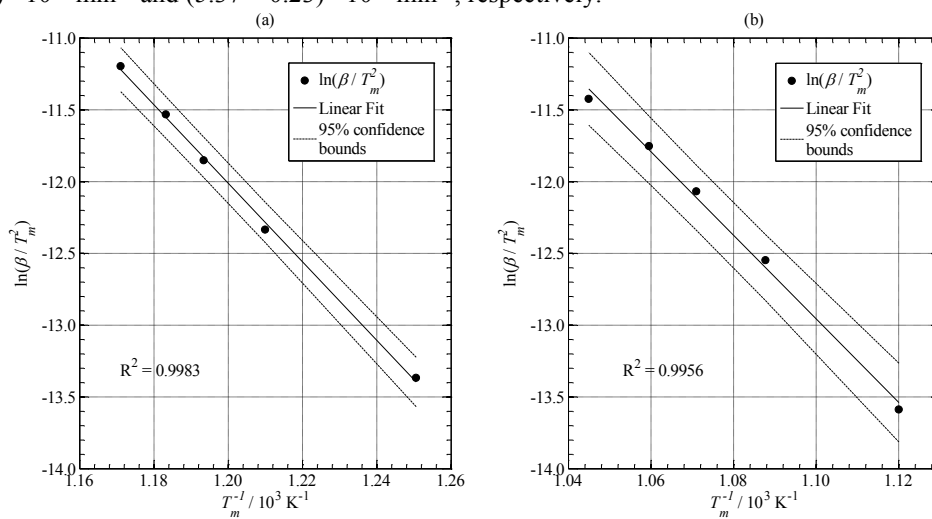

**FIGURE 2.** The Kissinger-type plot: (a) *tv*-layer dehydroxylation and (b) *cv*-layer dehydroxylation.

## CONCLUSIONS

The illite dehydroxylation was studied using differential scanning calorimetry in a dynamic argon atmosphere. The values of apparent activation energy and pre-exponential factor were determined using the Kissinger method, and the mechanism of each step of illite dehydroxylation was determined using the Avrami peak factor. The summary of the results is as follows:

- The calculated values of apparent activation energy and pre-exponential factor for *tv*-layer dehydroxylation are $(227 \pm 6)$ kJ mol$^{-1}$ and $(2.87 \pm 0.09) \times 10^{13}$ min$^{-1}$.
- The value of $n_A$ determined for the *tv*-layer dehydroxylation indicate 1D diffusion controlled reaction with instantaneous nucleation rate.
- The calculated values of $E_A$ and *A* for *cv*-layer dehydroxylation are $(242 \pm 10)$ kJ mol$^{-1}$ and $(5.37 \pm 0.23) \times 10^{12}$ min$^{-1}$.
- The value of the Avrami peak factor determined for this step indicate diffusion controlled growth of the new phase with zero or decreasing nucleation rate.

## ACKNOWLEDGMENTS

This work was supported by the grants UGA VII/1/2018, VII/16/2018 and by the Ministry of Education, Youth and Sports of the Czech Republic, the project number P105/12/G059.